# Prediction of Alpha-Particle-Immune Gate-All-Around Field-Effect Transistors (GAA-FET) Based SRAM Design

Albert Lu
Electrical Engineering
San Jose State University
San Jose, USA
albert.lu@sjsu.edu

Reza Arghavani
Sandia National Laboratories
Albuquerque, USA
rarghav@sandia.gov

Hiu Yung Wong
Electrical Engineering
San Jose State University
San Jose, USA
hiuyung.wong@sjsu.edu*

***Abstract*—In this paper, using 3D Technology Computer-Aided-Design (TCAD) simulations, we show that it is possible to design a static random-access memory (SRAM) using gate-all-around field-effect-transistor (GAA-FET) technology so that it is immune to single alpha particle radiation error. In other words, with the design, there will be no single-event upset (SEU) due to alpha particles. We first use *ab initio* calculations in PHITS to show that there is a maximum linear energy transfer (LET), $LET_{max}$, for the alpha particle in Si and $Si_xGe_{1-x}$. Based on that, by designing a sub-7nm GAA-FET-based SRAM with bottom dielectric isolation (BDI), we show that the SRAM does not flip even if the particle strike is in the worst-case scenario.**

## I. INTRODUCTION

Alpha particles are known to be a major source of radiation that can be emitted from the impurities within the packaging materials of semiconductor devices [1]. When these alpha particles strike a semiconductor device, they can generate electron-hole pairs that can then affect the voltages and currents of a device momentarily. More specifically, this can lead to soft errors in semiconductor devices, such as content flipping in static random-access memory (SRAM).

To further enhance performance and increase SRAM density, scaling of transistor nodes has continued. Recent advancements in transistor nodes have led to the introduction of GAA-FETs. GAA-FETs have the gate wrapped around all sides of the channel, which gives it better gate control and thus better electrostatics in the same footprint compared to planar and FinFET technologies. Substrate leakage has also been handled in different ways for GAA-FETs and FinFETs, such as by using substrate anti-punch-through doping or bottom dielectric isolation (BDI). BDI is created by placing an oxide layer to isolate the source/drain from the substrate. This is expected to help enhance the radiation hardness of an SRAM.

3D Technology Computer-Aided-Design (TCAD) is used to study semiconductor devices and has been used in the literature to study the radiation hardness of devices. TCAD is necessary for a detailed study of the device as well as many radiation strike parameters such as location, direction, and intensity. In physical experiments, it is highly impractical to obtain detailed data, such as the electrostatic potential or electron density, at every location in the device. This data can be useful for studying the effects of radiation strikes.

Many studies in the literature have performed radiation strike simulations on GAA-FETs and FinFETs [2]-[7][14]. However, many of them only study a single transistor or part of an SRAM configuration [4][5][7]. A few of them study a full 6T-SRAM configuration [2][3][6][14], which will also be studied in this work. It is important to simulate the SRAM as a whole in order to capture the effects of particle strikes that are layout-dependent. One such layout-dependent case is when a particle strikes through two transistors in a certain state that can cause both of them to turn on momentarily and thus more easily induce a bit flip. A few studies focused on the effects of GAA-FET with and without BDI [5]-[7]. It was shown that BDI makes the device more radiation-hard.

In this paper, TCAD Sentaurus is used. Sentaurus Device [8] and Sentaurus Process [9] are used to study the effects of radiation strikes on GAA-FET with and without BDI. This work studies the worst-case strike locations and some worst-case layout-dependent scenarios for an SRAM. It will be further shown that it is possible to design an SRAM using GAA-FET technology so that it is immune to single alpha particle radiation errors. *Ab initio* calculations in PHITS [10] will first be performed to show that there is a maximum linear energy transfer (LET), $LET_{max}$, for an alpha particle in silicon and various mole fractions of silicon germanium. Then, the SRAM will be constructed for both cases with and without BDI. Alpha particle strikes will then be conducted, and it will be shown that even in the worst-case scenarios, the GAA-FET-based SRAM with BDI will not have a bit flip even with LET equal to $LET_{max}$.

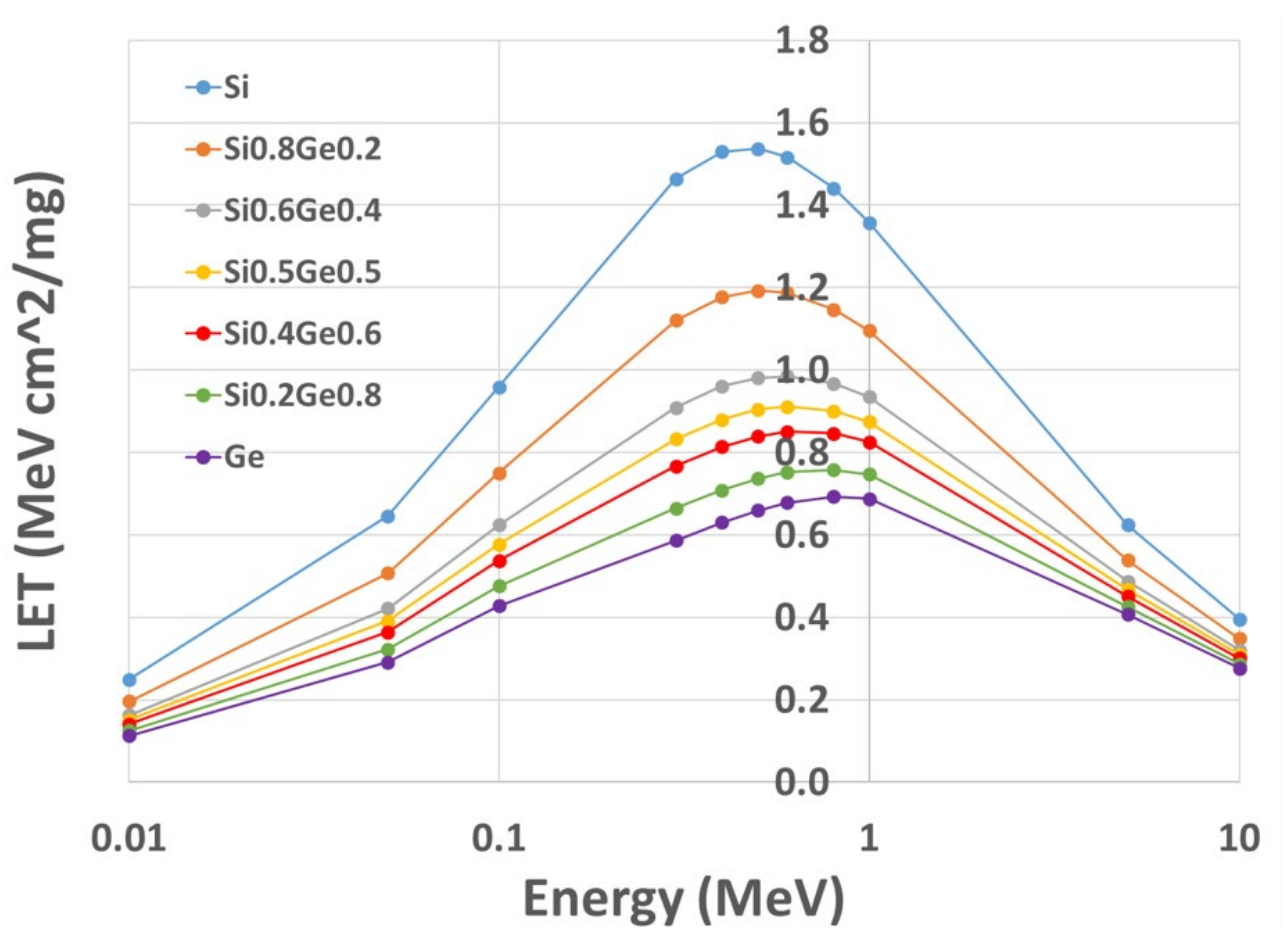

Fig. 1. Alpha particle LET vs. energy plot for $Si_xGe_{1-x}$ calculated using PHITS.

*Corresponding author: hiuyung.wong@sjsu.edu

| Parameter | Type 1 | Type 2 |
|---|---|---|
| Doping | No | Yes |
| Bottom Dielectric Isolation (BDI) | Yes | No |
| Inner Spacer | 7nm | 7nm |
| Oxide Thickness ($t_{ox}$) | 1nm | 1nm |
| $HfO_2$ Thickness ($t_{hk}$) | 2nm | 2nm |
| Oxide Dielectric Constant | 3.5 | 3.5 |
| $HfO_2$ Dielectric Constant | 20 | 20 |
| Silicon Sheet Thickness ($t_{si}$) | 5nm | 5nm |
| Sheet-to-Sheet Spacing ($t_{sus}$) | 10nm | 10nm |
| Gate Length (Gate Metal) ($L_G$) | 12nm | 12nm |
| Gate Length (Gate Metal + 2 $t_{high\ k}$) ($L_{G1}$) | 16nm | 16nm |
| Sheet Width (W) | 7nm | 7nm |
| N-Type Workfunction | 4.25 | 4.25 |
| P-Type Workfunction | 4.85 | 4.85 |

Fig. 2. Design parameters of the GAA-FETs used in this study. Their differences are highlighted. Type 1 is with BDI, and Type 2 is without BDI.

## II. Alpha Particle LET Calculation

The Particle and Heavy Ion Transport Code System (PHITS) is an *ab initio* tool that can be used to calculate the LETs of alpha particles [10]. It is a Monte-Carlo particle transport simulator that can record the LET of particles in a material. The approach to calculate LET using *ab initio* tools (PHITS or other tools) has been used previously for alpha particles [2] and even for neutrons [2][11][12], which demonstrates the robustness of the approach. Version 3.341 is used in this study.

Fig. 1 shows the extracted LET versus the energy of an alpha particle in silicon and silicon germanium with various mole fractions. It can be seen that the LET of the alpha particle in the various materials has a maximum peak at a specific particle energy. The maximum LET occurs when the alpha particle has an energy of 0.5 MeV striking into silicon, resulting in an $LET_{max}$ = 1.54 MeV·cm$^2$/mg. After a conversion factor from TCAD as discussed in [2], this $LET_{max}$ of 1.54 MeV·cm$^2$/mg is equivalent to 0.0144pC/μm and this will be used in TCAD simulations. If the SRAM can withstand this maximum LET, then it is possible for it to be immune to single alpha particle strikes.

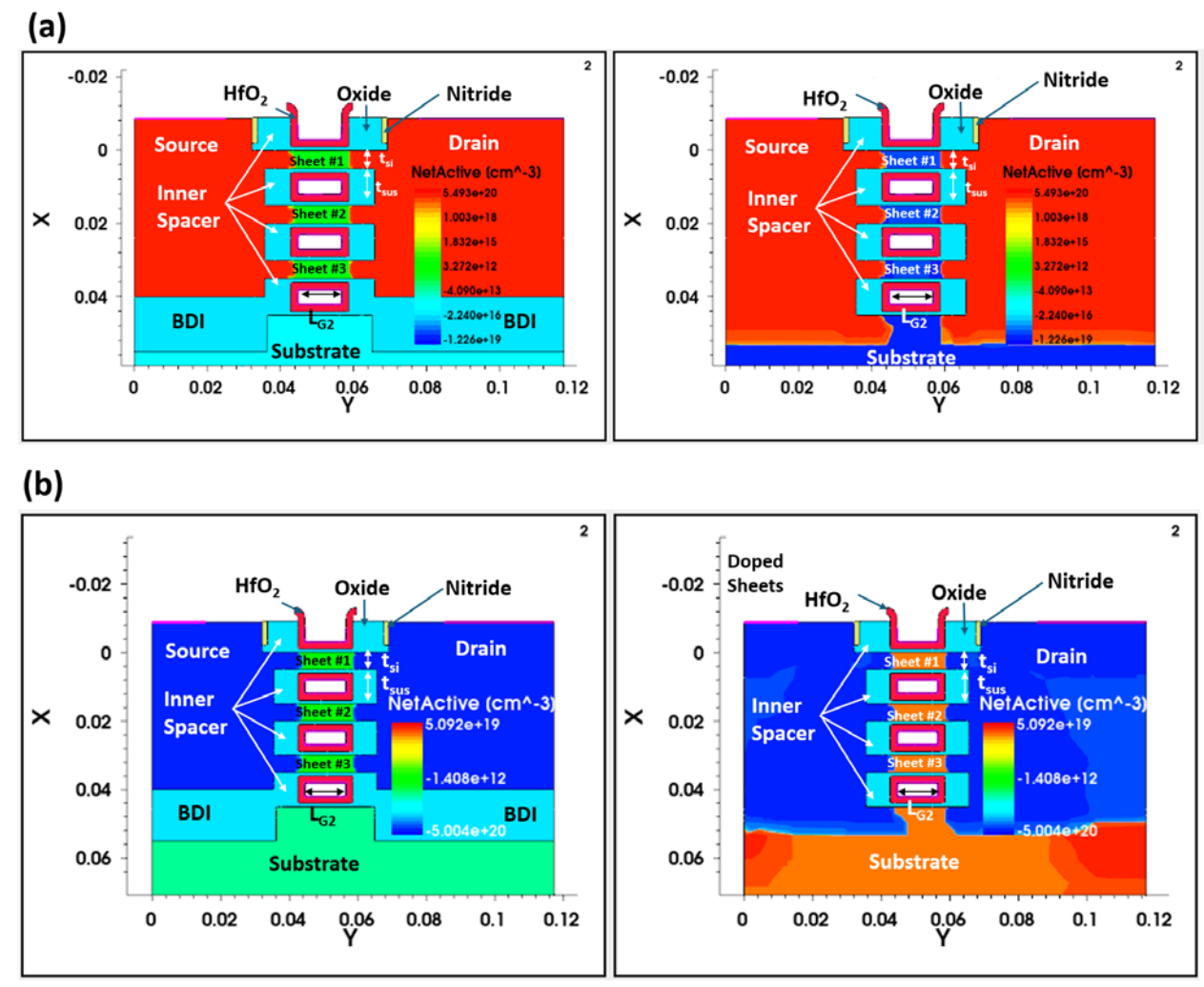

Fig. 3. Cross-section in the source/drain direction of GAA-FETs. (a): n-type. (b): p-type. Left: Type 1 with BDI. Right: Type 2 without BDI.

## III. TCAD Simulations And Results

### A. Transistor Design

Sentaurus Process was used to create the device structure from a layout. Ballistic mobility model, Philips Unified Mobility Model (PhuMob), Lombardi for surface scattering, impact ionization, SRH and Auger for recombination, and density gradient for quantum corrections are turned on in device simulations. Traps are also enabled at the Silicon/Oxide interface.

Two types of GAA-FET are studied. "Type 1" is a GAA-FET with BDI. "Type 2" is a GAA-FET without BDI, which requires a substrate punch-through stop implant (or halo implant) to avoid substrate leakage. Both types mimic the design rule of the sub-7nm GAA-FET in [13]. In this work, it should be noted that a "partial BDI" scheme is used, like in [5] and [6], where the BDI is only under the source/drain epitaxy. A "full BDI" scheme like in [7], where the BDI would extend across the source/drain and beneath the nanosheets, is not used in this study.

Fig. 2 shows the design parameters of the devices. Each GAA-FET has three sheets. It should be noted that the gate length, $L_G$, is 12nm without including the high-k dielectric on the sidewall of the inner spacer.

Fig. 3 shows the cross-section for both n-type and p-type devices along the source/drain and the definitions of various parameters. The sheets of both n-type and p-type GAA-FETs are silicon, but the p-type GAA-FET has $Si_{0.8}Ge_{0.2}$ source-drain epitaxy. The source/drain epitaxy doping is $5\times10^{20}$ cm$^{-3}$. The BDI is underneath only the source/drain as intended. It can also be seen that the sheet spacing is filled, from outermost to innermost, as oxide, $HfO_2$, and then metal, where the metal is removed and replaced by electrodes in the simulation. Fig. 4 shows the drain current–gate voltage ($I_D$-$V_G$) curves of the transistors. "Type 1" and "Type 2" have similar on-state currents and threshold voltages. Their subthreshold slopes are also similar.

### B. SRAM Design and Butterfly Curves

Fig. 5 shows the SRAM circuit. It is composed of 6 transistors. ACC1 and ACC2 are the access transistors. P1 and P2 are pull-up transistors. N1 and N2 are pull-down transistors. CH and CL designate the node voltages of the SRAM cell. BLB and BL are the bit lines. WL is the word line. In this figure, the SRAM is in a hold state and there is a cell on the same column being written.

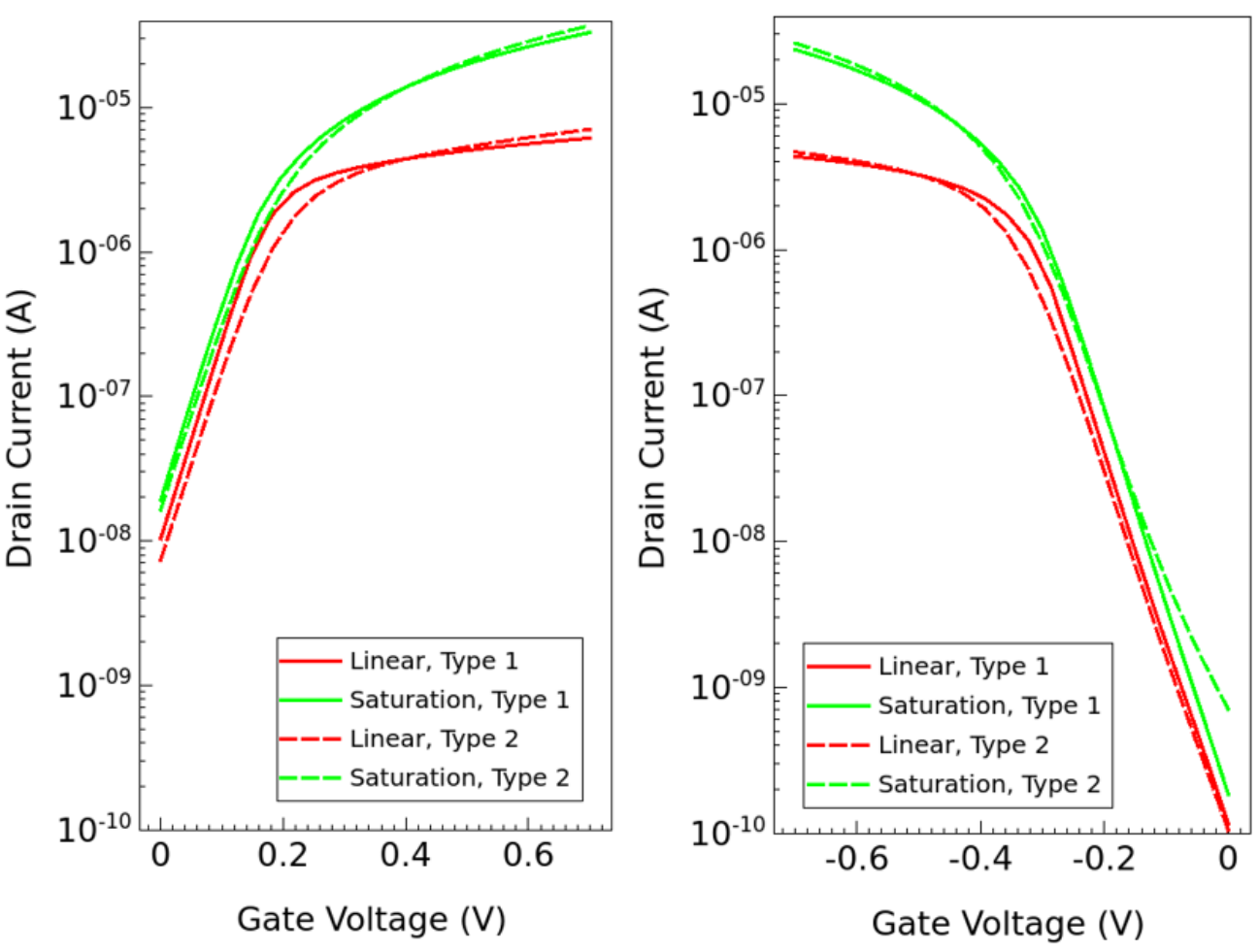

Fig. 4. Linear ($|V_D|$ = 50 mV) and saturation ($|V_D|$ = 0.7V) $I_D$-$V_G$ curves for the n-type (left) and p-type (right) devices.

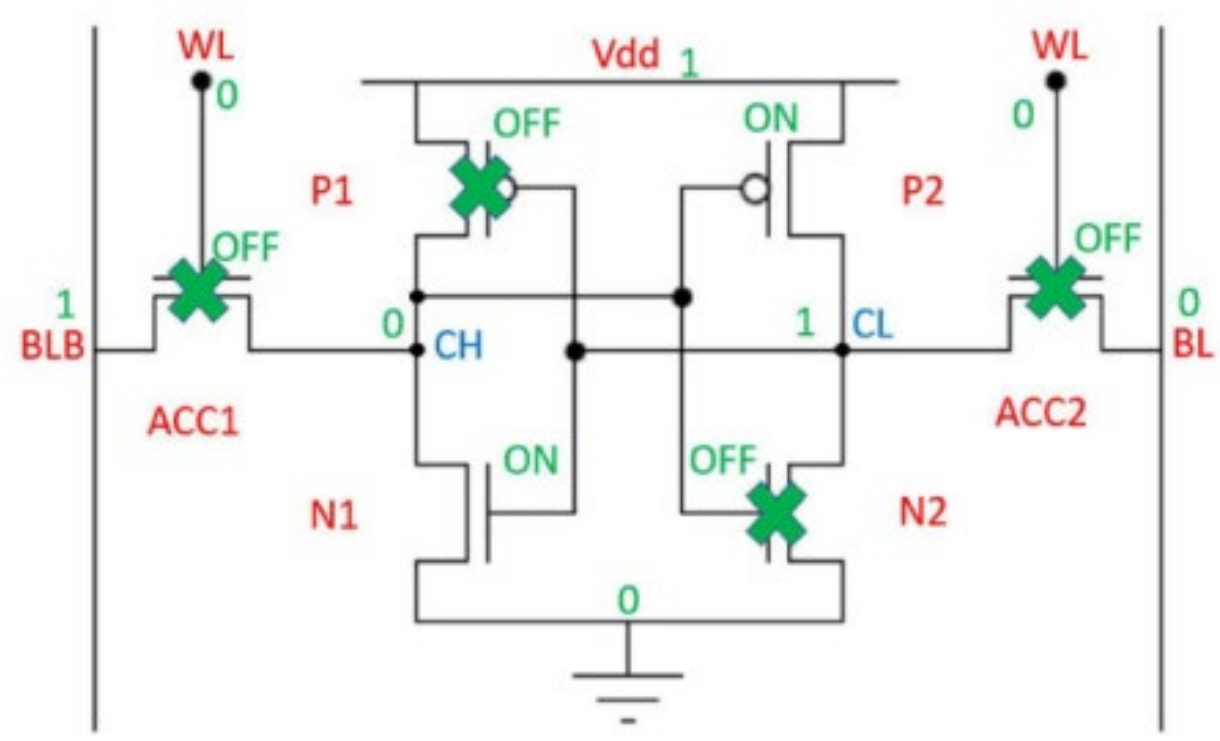

Fig. 5. Equivalent 6T-SRAM circuit used. The bias and ON/OFF conditions refer to the worst-case scenario during a radiation strike.

Fig. 6 shows SRAMs built with either "Type 1" or "Type 2" GAA-FETs. It was built in Sentaurus Process using the same 6T-SRAM GDS corresponding to the circuit in Fig. 5. It can be seen that "Type 1" has the BDI and thus the substrate is not connected to the sheets and source/drain epitaxy. There are two types of colors for the source/drain epitaxy. Red represents a high n-type doping, and blue represents a high p-type doping. It should also be noted that the n-type transistors have two fins, and the p-type transistors have one fin. For a proper SRAM design, the access transistors should be made weaker than the pull-down transistors. Since the access transistors are n-type and thus have two fins, one fin is turned off by setting its gate to always be 0 V. So, the access transistors are weaker than the pull-down transistors but are still designed to be stronger than the pull-up transistors, which are needed for a proper SRAM design.

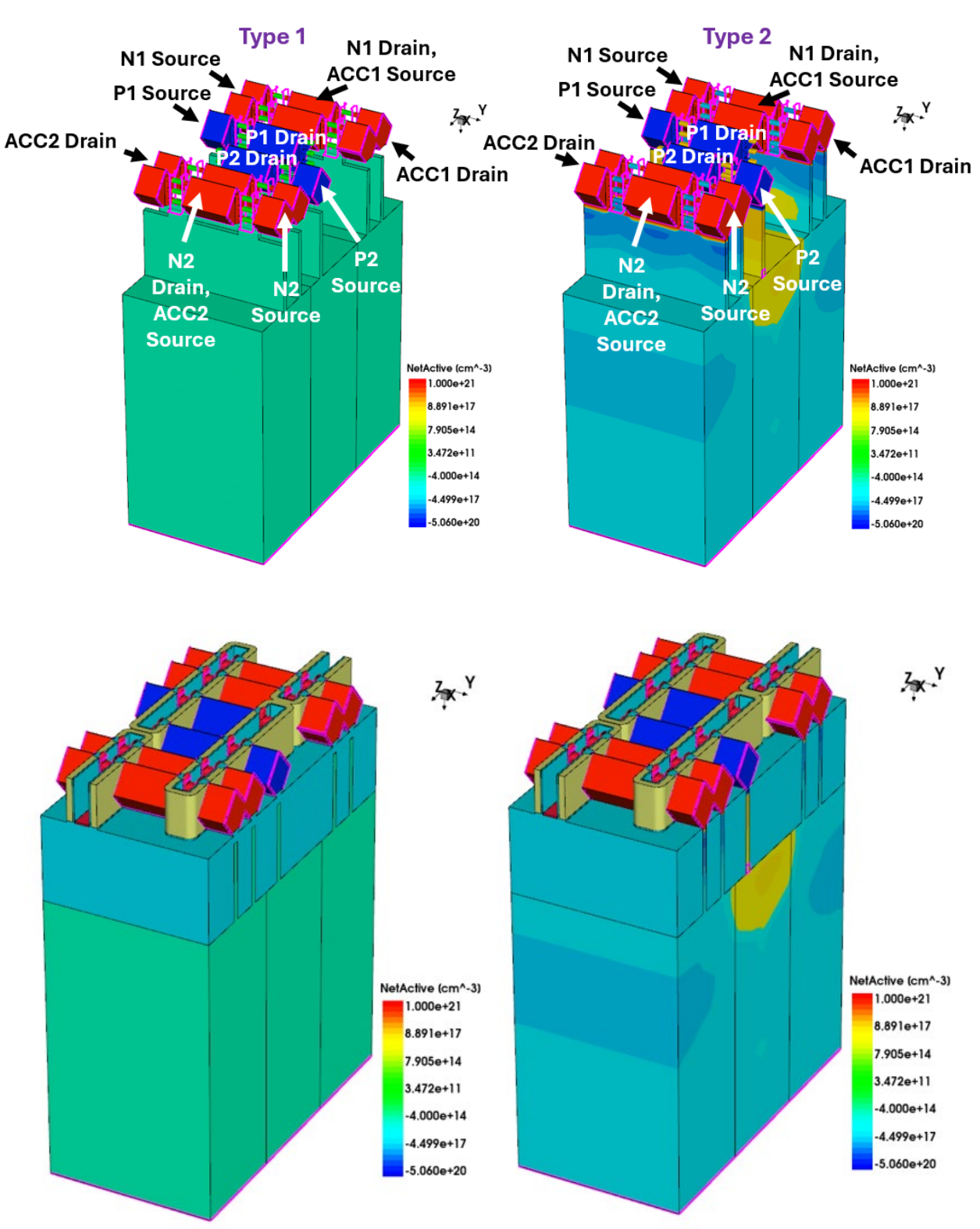

Fig. 6. SRAMs made of "Type 1" (left), and "Type 2" (right) GAA-FETs corresponding to the circuit in Fig. 5. For clarity, the top shows the structures without oxide, nitride, and $HfO_2$ layers. The bottom shows the structures with all layers enabled.

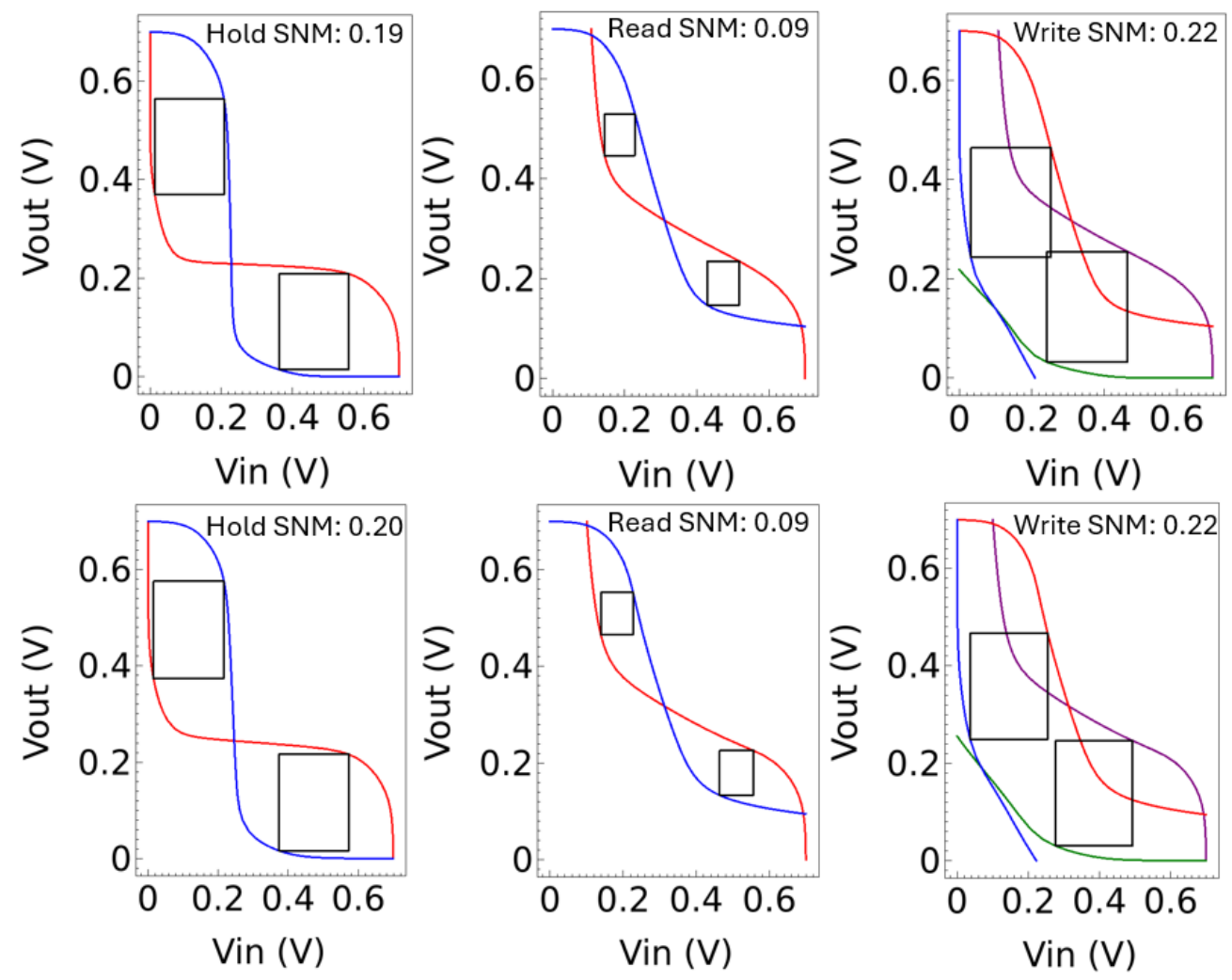

Fig. 7. Hold, read, and write butterfly curves of "Type 1" (top row) and "Type 2" (bottom row) SRAMs. The static-noise-margins (SNM) are given in V.

Fig. 7 shows the hold, read, and write butterfly curves as well as their respective static noise margins (SNMs) of both SRAMs, and they have similar SNMs. This is expected as their transistors have similar $I_{on}$ and $I_{off}$. So, they *have the same stability from the circuit design perspective*. However, they might not have the same radiation hardness, which depends on transistor design.

## C. Radiation Hardness Simulation

Based on [14], radiation strikes are performed at the most vulnerable positions and conditions. The LET of the strikes is then varied to understand when a bit flip occurs. Note again that the $LET_{max}$ of an alpha particle in Si was determined earlier to be 0.0144pC/μm, which sets a practical restriction on what LET values are useful to consider. The strikes follow a Gaussian distribution with a peak value of 50 ps and a standard deviation of 2 ps. Before the radiation strikes, the SRAM is first initialized into the state shown in Fig. 5.

One vulnerable position is identified in Fig. 5, during which a particle strikes both the off-state access and pull-down transistors (N2 and ACC2) while the transistor being struck is in a hold state and there is a cell on the same column being written. When N2 and ACC2 are turned on at the same time, this may cause node CL to discharge more easily. This demonstrates the importance of considering the effect that the layout has on radiation hardness. Fig. 8 shows the results of a “channel” strike, where the strike is at the middle sheet such that it goes through N2 and ACC2 simultaneously. It can be seen that “Type 1” (with BDI) and “Type 2” (without BDI) flipped at about the same LET. Therefore, there is not much difference in radiation hardness for this case.

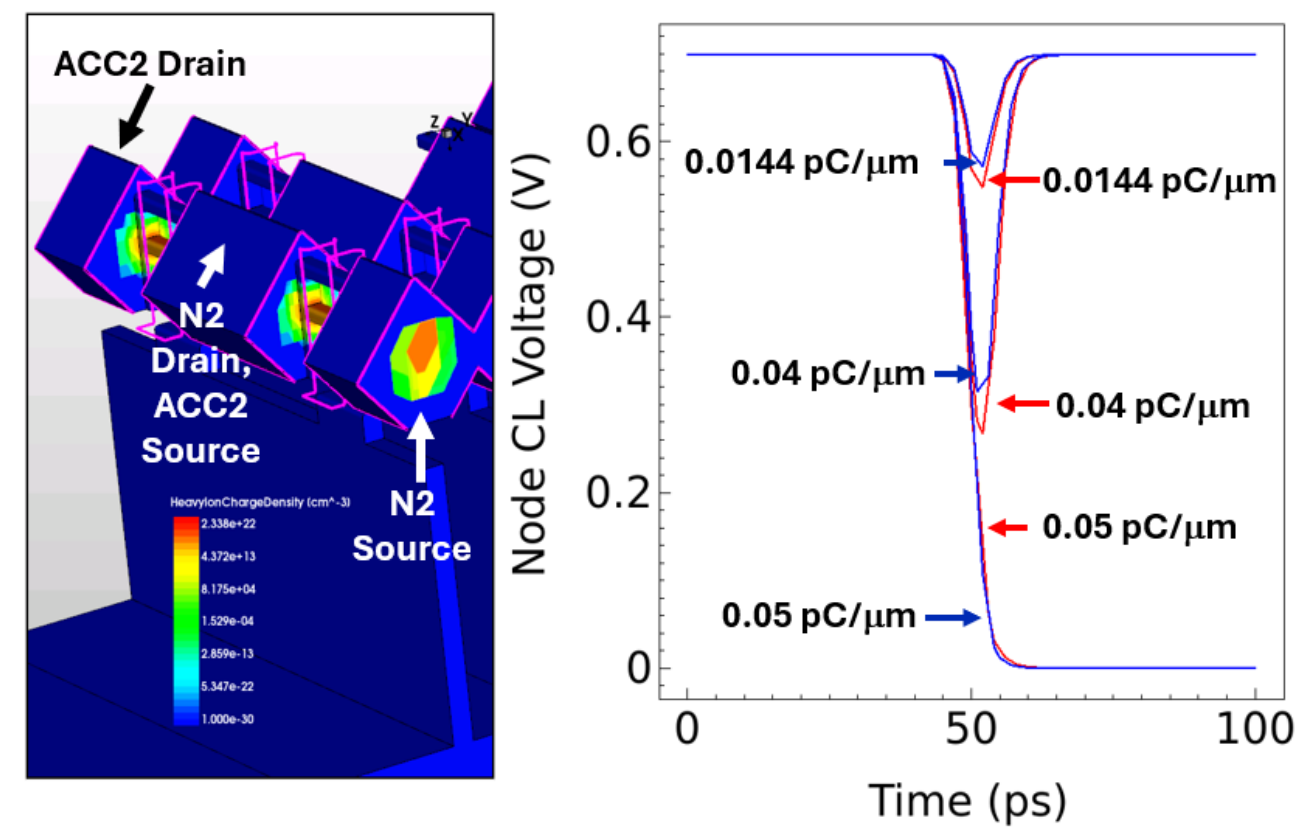

Fig. 8. Left: Channel strike from N2 to ACC2. Right: CL node voltage as a function of time. Red is "Type 1". Blue is "Type 2".

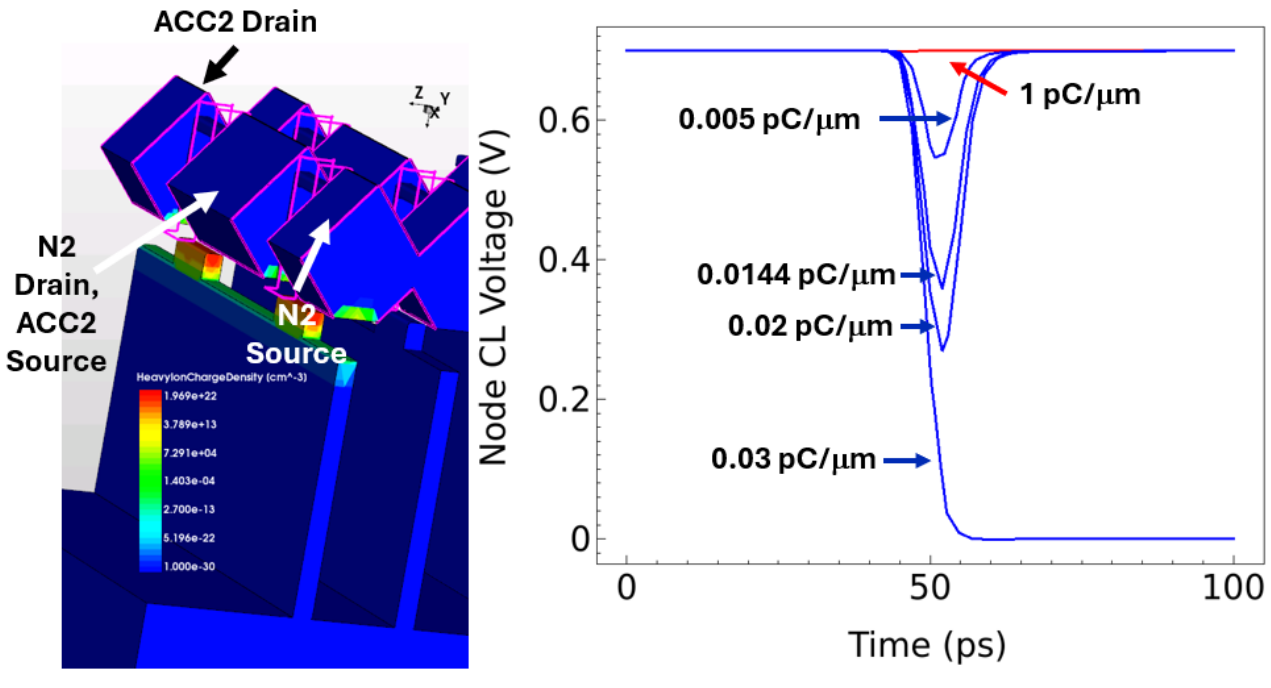

Fig. 9. Left: Substrate strike at N2 from N2 to ACC2. Right: CL node voltage as a function of time. Red is “Type 1”. Blue is “Type 2”.

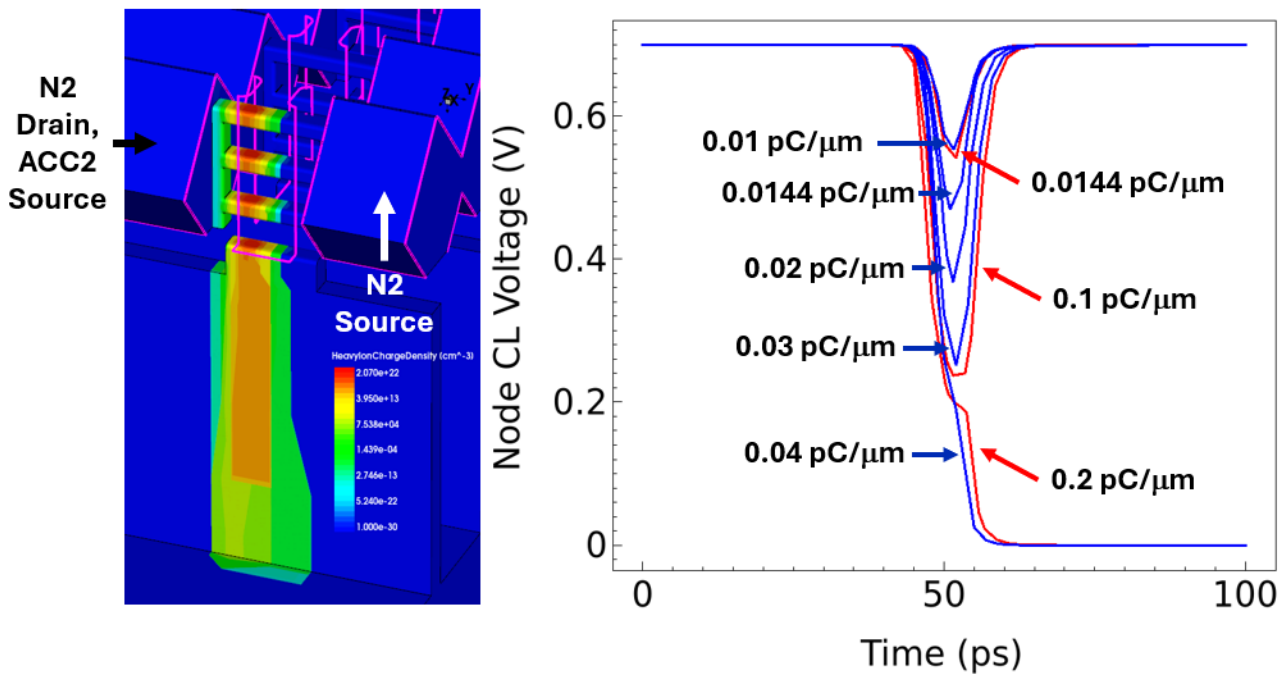

Fig. 10. Left: Vertical strike at N2 through 3 nanosheets (10.9nm from the drain epi. 10.9nm is tested to be the most sensitive location.). Right: CL node voltage as a function of time. Red is “Type 1”. Blue is “Type 2”.

Fig. 9 shows the result of a “substrate” strike, where the strike location is from N2 to ACC2 but along the substrate under the nanosheets. “Type 2” already flips at an LET as low as $2.08LET_{max}$ but “Type 1” does not flip even at $69LET_{max}$ because there is no semiconductor along the striking path. Thus, this does not affect its SRAM node voltage. This demonstrates how BDI effectively increases the radiation hardness of the device by reducing the number of sensitive regions of the device.

Another sensitive location is near the drain of N2 when the strike goes vertically through three nanosheets. This is a “top” strike, and it is positioned 10.9nm from the drain epitaxy. Fig. 10 shows that while “Type 2” flips at 2.78$LET_{max}$, “Type 1” does not flip until 13.88$LET_{max}$. Other “top” strikes further from the drain of N2 were found to be even less sensitive.

## IV. Conclusions

Using *ab initio* and rigorous TCAD simulation, the maximum LET of an alpha particle in silicon and different mole fractions of silicon germanium are extracted, and a GAA-FET-based SRAM is created. Two types of SRAMs (with BDI and without BDI) are considered. Radiation strikes were then conducted in a few vulnerable locations, and it was found that the “Type 1” case was significantly more radiation-hard. “Type 1” flips only if the LET goes past the $LET_{max}$. Thus, we predict that it is possible to design a 6T-SRAM cell that is immune to alpha-particle-induced SEU especially by using the novel GAA-FET technology with BDI.

## Acknowledgment

Sandia National Laboratories is a multimission laboratory managed and operated by National Technology & Engineering Solutions of Sandia, LLC, a wholly owned subsidiary of Honeywell International Inc., for the U.S. Department of Energy’s National Nuclear Security Administration under contract DE-NA0003525.

## References

[1] R. C. Baumann, "Soft errors in advanced semiconductor devices-part I: the three radiation sources," in IEEE Transactions on Device and Materials Reliability, vol. 1, no. 1, pp. 17-22, March 2001, doi: 10.1109/7298.946456.

[2] J. Saltin et al., "FinFET and Nanowire SRAM Radiation Hardness Studies using Ab initio TCAD Simulation Framework," arXiv preprint arXiv:2202.13769, 2022. [Online]. Available: https://arxiv.org/abs/2202.13769

[3] Khoa Huynh, Johan Saltin, Jin-Woo Han, Meyya Meyyappan and Hiu Yung Wong, "Study of Layout Dependent Radiation Hardness of FinFET SRAM using Full Domain 3D TCAD Simulation," 2019 IEEE SOI-3D-Subthreshold Microelectronics Technology Unified Conference (S3S), San Jose, CA, USA, 2019, pp. 1-3, doi: 10.1109/S3S46989.2019.9320706.

[4] J. Kim, J. -S. Lee, J. -W. Han and M. Meyyappan, "Single-Event Transient in FinFETs and Nanosheet FETs," in IEEE Electron Device Letters, vol. 39, no. 12, pp. 1840-1843, Dec. 2018, doi: 10.1109/LED.2018.2877882.

[5] X. -T. Zheng and V. P. -H. Hu, "Improved Radiation Hardness for Nanosheet FETs with Partial Bottom Dielectric Isolation," *2023 Silicon Nanoelectronics Workshop (SNW)*, Kyoto, Japan, 2023, pp. 75-76, doi: 10.23919/SNW57900.2023.10183963.

[6] M. Bang *et al*., "Mitigation of Single Event Upset Effects in Nanosheet FET 6T SRAM Cell," in *IEEE Access*, vol. 12, pp. 130347-130355, 2024, doi: 10.1109/ACCESS.2024.3457750.

[7] G. Choi and J. Jeon, “Radiation effects on multi-channel Forksheet-FET and Nanosheet-FET considering the bottom dielectric isolation scheme,” Nuclear Engineering and Technology, vol. 56, no. 11, pp. 4679–4687, Nov. 2024, doi: https://doi.org/10.1016/j.net.2024.06.031.

[8] Sentaurus™ Device User Guide Version V-2024.03, March 2024.

[9] Sentaurus™ Process User Guide Version V-2024.03, March 2024.

[10] K. Niita et al., Particle and Heavy Ion Transport Code System; PHITS, Radiat. Meas. 41(9-10), 1080-1090,2006

[11] X. Jin *et al*., "Simulation and Experiment in Neutron Induced Single Event Effects in SRAM," *2017 17th European Conference on Radiation and Its Effects on Components and Systems (RADECS)*, Geneva, Switzerland, 2017, pp. 1-5, doi: 10.1109/RADECS.2017.8696259.

[12] C. Peng *et al*., "Incorporation of Secondary-Ion Information and TCAD Simulation for Atmospheric Neutron Soft-Error-Rate Prediction in SRAMs," in *IEEE Transactions on Nuclear Science*, vol. 66, no. 10, pp. 2170-2178, Oct. 2019, doi: 10.1109/TNS.2019.2938988.

[13] N. Loubet et al., "Stacked nanosheet gate-all-around transistor to enable scaling beyond FinFET," 2017 VLSI, Kyoto, Japan, 2017, pp. T230-T231.

[14] A. Elwailly et al., "Radiation Hardness Study of LG = 20 nm FinFET and Nanowire SRAM Through TCAD Simulation," in IEEE TED, vol. 68, no. 5, 2021.